# Paired teaching for faculty professional development in teaching

Jared B. Stang* and Linda E. Strubbe^

Department of Physics and Astronomy, University of British Columbia, 6224 Agricultural Road, Vancouver, BC, Canada V6T 1Z1

*jared@phas.ubc.ca
^linda@phas.ubc.ca

**Abstract:** Paired (or co-)teaching is an arrangement in which two faculty are collaboratively responsible for all aspects of teaching a course. By pairing an instructor experienced in research-based instructional strategies (RBIS) with an instructor with little or no experience in RBIS, paired teaching can be used to promote the adoption of RBIS. Using data from post-course interviews with the novice instructors of four such arrangements, we seek to describe factors that make for effective professional development in teaching via paired teaching. We suggest that the novice instructor's approach to the paired teaching and their previous teaching experience are two aspects which mediate their learning about teaching. Additionally, the structure of the pair-taught course and the sequence of teaching assignments for the novice instructor both likely play roles in facilitating the adoption of RBIS by novice instructors. We discuss these results within the framework of cognitive apprenticeship.

# Paired teaching for faculty professional development in teaching

**Abstract:** Paired (or co-)teaching is an arrangement in which two faculty are collaboratively responsible for all aspects of teaching a course. By pairing an instructor experienced in research-based instructional strategies (RBIS) with an instructor with little or no experience in RBIS, paired teaching can be used to promote the adoption of RBIS. Using data from post-course interviews with the novice instructors of four such arrangements, we seek to describe factors that make for effective professional development in teaching via paired teaching. We suggest that the novice instructor's approach to the paired teaching and their previous teaching experience are two aspects which mediate their learning about teaching. Additionally, the structure of the pair-taught course and the sequence of teaching assignments for the novice instructor both likely play roles in facilitating the adoption of RBIS by novice instructors. We discuss these results within the framework of cognitive apprenticeship.

## I. Introduction

There is strong evidence that the use of active learning strategies (e.g., in-class worksheets, peer instruction, group problem solving), compared with pure lecture, increases student performance and reduces failure rates in science, engineering, and mathematics classrooms [1]. Unfortunately, system-wide adoption of so-called "research-based instructional strategies" (RBIS) has been slow. For example, although the vast majority of physics faculty in the United States have familiarity with one or more RBIS, less than half reported using at least one RBIS [2] and approximately 1/3 of faculty discontinued use of an RBIS after trying it [3]. Furthermore, it is common for faculty to make significant modifications to the implementation of RBIS which may hamper their effectiveness [2]. In order to promote the (proper) use of RBIS, more support of faculty in adopting these methods is needed.

Paired (or co-)teaching, in which two faculty are collaboratively responsible for all aspects of teaching a course, has been suggested as an effective method for the dissemination of RBIS [4]. In contrast to traditional dissemination strategies that employ a transmissionist approach to learning (such as talks), paired teaching is a long-term professional development experience with built-in feedback mechanisms—components that have been identified as characterizing successful change strategies [5].

Due to the efforts of the Carl Wieman Science Education Initiative (CWSEI) [6, 7], significant expertise in evidence-based teaching exists in the Department of Physics and Astronomy at UBC [8]. In this extension to the work of the CWSEI, we are investigating the use of paired teaching in leveraging this expertise to promote further instructional change in our department.

In this paper, we describe preliminary results in this project, focusing on the effectiveness of paired teaching for faculty professional development in teaching. Our research question is:

> *What factors contribute to effective development in teaching via paired teaching?*

By studying four cases in which a faculty member not experienced in RBIS (the "novice") is paired with a faculty member experienced in RBIS (the "expert"), we aim to understand these factors and, ultimately, provide recommendations for units interested in adopting this dissemination strategy.

## II. Context & Method

The paired teaching arrangements took place in courses 1 and 2, a calculus-based first-year physics sequence which serves mostly non-physics majors. There were three different class sections in both courses 1 and 2. Thus, the paired teaching occurred within a "team-teaching" environment, as one section was pair-taught while the other two sections were taught by individual instructors. Assessments across the sections were common and the teams of instructors met weekly. The class size for the pair-taught sections of course 1 varied between 240-280 students, while the pair-taught section of course 2 had 100 students. The lecture portion of course 1 has been transformed to an active structure due to CWSEI activities [7], and this structure is carried from year to year. Course 2 has also been transformed, and underwent further development in the 2015 year in an initiative undertaken by expert Instructor Z. Observations in course 1 and 2 in 2015 reveal extensive use of active learning techniques in each class, with about 25% of the time spent in the lecture mode (delivery of content) and most of the remaining time spent in interactive learning modes, such as clicker questions or group activities (worksheets).

In 2015, both paired teaching arrangements had the support of a science education specialist (SES, author JBS). The SES performed classroom observations and informal interviews with students, providing this feedback to the instructor pairs weekly.

We focus on the professional development of the four novice instructors, whom we refer to as Instructors A, B, C, and D. To distinguish the expert partners clearly, we refer to them as expert Instructors Y and Z. In order to protect the identity of the instructors, we have chosen to refer to the instructors using gender-neutral pronouns such as "they" and "them."

Post-course interviews with the novice instructors were conducted by the author JBS. (The interview protocol is reproduced in Appendix A.) For Instructors A and B, these interviews took place 6 to 18 months after pair-teaching, while for Instructors C and D, the interviews took place within a few weeks of the end of their pair-taught course. These interviews were transcribed and analyzed for evidence pertaining to: 1. The relevant "input" factors that characterize paired teaching arrangements; 2. The novice instructors learning about teaching; 3. Connections between the input factors and faculty outcomes. After consulting with colleagues who are expert in qualitative education research, the authors used an open coding approach to collaborate in an iterative process of independent coding and comparing to develop the themes present in the transcripts [9].

In terms of input factors, the major themes that emerged from the interview data were about the attitudes of the participants, the relationship of the paired instructors, and the types of interactions that the pairs had. The main categories for ideas concerning the novice instructor learning about teaching included teaching skills, teaching strategies, affective statements about RBIS, and reflections on teaching. Connections between the input factors and faculty outcomes were derived both from specific comments in interviews and from circumstantial evidence.

There was agreement between the authors on all of the major conclusions drawn, and the few disagreements on more minor issues were resolved through discussion. In addition, some objective information (such as the teaching backgrounds of instructors) is reported from other sources.

To evaluate learning about teaching, in addition to the evidence present in the interview transcripts, we also collected data about which teaching strategies the instructors are using. These data were collected using the Teaching Practices Inventory [10] (a self-report tool for characterizing the teaching practices used in science courses) and structured in-class observations [11]. We propose that strong evidence of learning would be the transfer of teaching techniques to an antagonistic scenario (e.g., a course which has no history of using RBIS). Moderate evidence of learning might include the use of RBIS while teaching the same course again individually or a positive shift in professed attitudes towards RBIS. Using RBIS *while* pair-teaching is only weak evidence; as described below, the existing course structure in our examples means that novice instructors are very likely to teach in a reformed style while pair-teaching (but that does not guarantee they will teach this way in other future courses). For this preliminary report, our data set is limited to results about the use of RBIS while pair-teaching (structured observations for Instructors C and D), and results about the use of RBIS in teaching the same course again individually (TPI results for Instructors A and B in course 1).

We use a case study design for this exploratory research into paired teaching, with the goals of developing hypotheses about which factors are important for effective professional development, and of uncovering directions for further inquiry. This approach is appropriate given the complex social nature of paired teaching, our lack of control over the behavioural events that occur within paired teaching (in contrast to the situation in a controlled experiment), and our focus on paired teaching as a contemporary event [12]. We undertake to ensure internal validity by using multiple sources of evidence (e.g., interviews, the TPI), and by placing our study within the existing framework of cognitive apprenticeship (section III B). Although the nature of this approach means that our conclusions may not generalize, the case study design allows us to use this rich dataset to make preliminary recommendations, and to inform our future research study design.

## III. Results & Discussion

### A. Factors in paired teaching arrangements

In this subsection, we describe some of the factors that characterize paired teaching. These are summarized in Table I; we elaborate on some of the factors here.

As summarized in Table I, the approach and attitude that each novice instructor took toward paired teaching differed. Instructors A and B took a developmental approach to this experience. Instructor B explained that "really all my interactions mostly are attempting to get feedback [from the expert].'' Instructor C took a more tempered approach, acknowledging that "most of the things weren't new to me,'' but still seeking feedback from expert Instructor Y both in person and through email. In the post-semester interview, Instructor D discussed paired teaching mostly in terms of creating a better product for the students rather than in terms of Instructor D's own professional development. Instructor D described paired teaching as "a super teaching tool'' and referred to the other instructors of course 2 also as "co-teachers'' (our local term for paired teaching partners), indicating a difference in Instructor D's conceptualization of the paired teaching relative to those of Instructors A, B, and C.

An additional factor that differed between the cases was the support of the SES. For example, in course 1 in 2015, Instructor C and the SES developed an observation protocol, based on CWSEI work [13], to document student engagement with the in-class worksheets.

### B. The factors as aspects of the cognitive apprenticeship

Following the work of [4, 13], we consider our department's paired teaching arrangements in the framework of the cognitive apprenticeship instructional model, which aims to make the strategies and heuristics that experts use explicit. Collins et al. describe six aspects of the cognitive apprenticeship [14]: (1) *modeling* of the expert strategies, (2) *coaching* by the expert, (3) *scaffolding* for the novice, (4) *articulation* of the novice's thinking, (5) *reflection* by the novice, and (6) *exploration* of novel tasks or situations by the novice.

The pair-teachers alternated leading the teaching in different parts of the course (in some cases alternating throughout each lecture, in other cases alternating the first half and second half of the semester), providing opportunities for modeling, coaching, and some exploration. Scaffolding for using RBIS is provided by the course structure: Each course has been previously transformed to active learning, and course materials are carried forward from year to year. Pair-teacher meetings were informal and occasional (although weekly meetings for the whole instructor team did take place): these provide some space for coaching, articulation, and reflection (though perhaps not as much as dedicated weekly meetings would have). The SES observing and providing feedback to the pair is another part of coaching. Having novice instructors teach the same or different courses on their own in the future provides opportunity for exploration.

### C. The novice instructors learning about teaching

In this subsection, we describe the outcomes of paired teaching for each of the four novice instructors in turn.

Instructor A described paired teaching as "vital'' to their development as an instructor, and provided many examples of both specific teaching skills they learned (regarding, for example, lecture preparation, crowd management, and the ability to adapt to the students' needs in class) and a higher level approach to teaching (such as discussing the importance of active learning). An important theme that emerged was the development of overall confidence in teaching. Furthermore, through the experience, Instructor A developed an interest in the research basis of teaching techniques. "I didn't really expect to be that interested in the why of the questions.'' Subsequent to pair-teaching, Instructor A taught the same course again individually. Instructor A's TPI results for their most recent semester of teaching course 1 show a continued use of the evidence-based techniques (such as in-class problem solving and pre-reading with online quizzes [15]) that were used while pair-teaching.

Instructor B repeatedly referred to pair-teaching as an "apprenticeship model'' and with positive affect, indicating the importance of the experience for their development as an instructor. They identified several specific teaching skills they learned, including the need for adaptation while teaching. Since pair-teaching, Instructor B has taught in a variety of different situations, including an online course, a small cohort-based program, and in course 1 again. Instructor B's TPI results for their most recent semester of teaching course 1 show that they have continued using the evidence-based techniques used while pair-teaching in course 1. Furthermore, they are an active member in the department's physics education research group and have undertaken education research projects in collaboration with expert Instructor Y.

Instructor C identified learning several concrete teaching skills, including pacing and adaptation, and was generally positive about the use of RBIs. "I can't be argumentative about the use of classical lecture versus more interactive class [sic].'' They reflected about topics such as the overall course structure ("I wouldn't change... the balance of lecturing versus worksheets and things like that'') and the role of the instructor (as "being able to react and interact with rather than ... just delivering content up front''). In addition to expressing that they would teach course 1 "exactly the same,'' at several points in the interview, Instructor C described their plan to transfer the approach to their junior level course. "For the upper level class… I will try to see if I can develop guided worksheets" in order to "try and let them work things out more directly with their own brains on worksheets," in the style of a recent upper-division course transformation [16]. Structured observations of the pair-taught class showed that there were no qualitative differences in the frequency or length of use of evidence-based teaching strategies between Instructor C and expert Instructor Y.

Instructor D described a changing perspective in "thinking a little bit more like a student as opposed to just thinking like a lecturer in the traditional sense." However, they expressed some reservations about the course content (or the lack of content covered) and seemed to conflate the addition of active learning techniques with removing challenging course content. "The other thing that I'd still like to learn is… the blending of slightly more challenging aspects with still this way of being very interactive." Observations of the pair-taught class showed that there were no qualitative differences in the frequency or length of use of evidence-based teaching strategies between Instructor D and expert Instructor Z.

In summary, Instructors A and B appear to have developed a large variety of skills and pedagogical knowledge, which they have subsequently transferred to their later teaching assignments. Instructor C showed strong evidence of buy-in to the techniques used while pair-teaching and described concrete plans to transfer this pedagogical knowledge to a different context. Instructor D described some shift towards a more student-centered attitude, but was restrained in their buy-in for the techniques. While pair-teaching, Instructors C and D taught in a manner consistent with existing (reformed) course structures.

### D. Factors influencing the development of the novice instructors

In this subsection, we offer ideas on how the factors characterizing paired teaching (section III A) may have contributed to different learning outcomes for the novice instructors (section III C), including how they fit into the cognitive apprenticeship framework.

Based on our investigation of the novice instructors in these arrangements, we suggest that an important consideration in the success of paired teaching, which likely mediates the amount of pedagogical learning that occurs, is the approach and perception of the novice instructor. Instructors A and B took up their roles as novices with gusto: they went in with the deliberate intention of learning about teaching from their expert counterpart. These instructors explicitly connected observing the expert in the classroom with developing their own teaching practices. Instructors A and B also refer to the importance of the expert feedback in their situations. As Instructor A summarizes, "The most valuable [interaction] was actually me sitting in class… that was I think at least 50% of it. And then the other 50% came from both the discussions afterward and the feedback that I got when I was teaching." For Instructors A and B, it appears that taking a developmental approach meant that they were able to take advantage of both observing the expert and receiving feedback from them, resulting in their learning both pedagogical skills and knowledge. In contrast to these, Instructor D makes no direct statements connecting their development in teaching to expert Instructor Z's practices. As discussed above, they focused not on professional development in teaching but on creating a good product for the students. Although Instructor D appeared to teach in a reformed way during the paired-teaching experience (as dictated by the course structure), there was comparatively little evidence in their interview as to any internalized change. How Instructor D approached the paired teaching may have reduced their ability to learn from the professional development experience. In the framework of cognitive apprenticeship, the approach of the novice instructor to observing and receiving feedback would be important for maximizing the benefits of modeling and coaching.

We further suggest that what the novice instructors learn likely depends on their previous teaching experience. Instructor A, with less than one year of teaching experience and no prior experience with active learning techniques, discussed learning many basic skills (such as, "How does a clicker work in practice?") that Instructors C and D, with 10 years of experience, did not mention. Overall, the relatively less experienced novice instructors (A and B) reported learning more skills than the relatively more experienced novice instructors (C and D). The level of previous teaching experience could also affect instructors' openness to coaching.

The support and coaching from the SES also appears to have been important for Instructor C's developing attitude towards in-class activities. Based on their experiences this semester—which included the development of a worksheet engagement observation scheme—they conclude that "there is no doubt that they [worksheets] improve engagement."

The structure of the course also seems likely to be an important factor in paired teaching. In each case, the course structure was established and prior materials existed, offering scaffolding to the novice instructors for beginning to use RBIS during the pair-taught course.

The sequence of teaching assignments for the novice instructors may play a role in providing practice and in shaping intentions towards future instruction. Instructors A and B went on to teach the same (or similar) courses after pair-teaching, giving them the opportunity to put into practice the techniques that they learned. Both instructors continued to use many of these practices. Teaching an upper division course at the same time—and being scheduled for it next year—provided Instructor C a concrete example in which to speculate about transferring the teaching approach to a new situation. This fits into the framework of cognitive apprenticeship as an opportunity for exploration.

The relationship between the novice and expert instructors was identified by all novice instructors as important for the success of paired teaching. Instructor B observed that "compatibility really makes a big difference when you're doing this kind of work" while Instructor D opined, "I can also see that if the teachers don't get along that it can be a total disaster." Building a positive relationship—which, fortunately, all four of these cases were able to do—may be a necessary condition for positive outcomes. The relationship between the instructors is likely a key component for facilitating most aspects of the cognitive apprenticeship, especially coaching, articulation, and reflection.

### IV. Summary & Conclusions

Reporting on four paired teaching arrangements, we have described factors which characterize such arrangements and the learning outcomes that occurred for the novice instructors, and we have speculated on ways in which this learning may be influenced by the input factors.

We suggest that the approach of the novice instructor to pair-teaching likely affects their ability to learn about teaching: An instructor who approaches paired teaching with the intent to learn will likely get more out of it than an instructor who does not perceive the arrangement as professional development. Additionally, the previous teaching experience of the novice instructor may influence what they learn. In terms of the larger context, the structure of the pair-taught course and the sequence of teaching assignments for the novice instructor both likely play roles in facilitating the adoption of RBIS by novice instructors. Strategic teaching assignments may make the immediate use of new techniques and the subsequent transfer of these to new contexts more likely. The cognitive apprenticeship model offers one framework for understanding these results: effective coaching (through productive attitudes and positive relationships) appears to be among the most important aspects for the teaching pairs in this study.

Based on the preliminary results presented here, we may provide speculative recommendations for units considering paired teaching as a framework for the dissemination of RBIS:

- Ask instructors to volunteer (or even apply) to pair-teach
- Place teaching pairs in courses where interactive materials already exist
- Carefully map out future teaching assignments for pair-instructors
- Hold an orientation for teaching pairs to clarify goals and expectations, and support the development of a positive professional relationship

These results align with the promise described in [4] for using paired teaching to support the adoption of RBIS. Of course, the story of Instructors A-D is not yet complete, and some of the most important evidence is yet to come as they move on to different teaching contexts. Further work will continue to evaluate both these and future paired teaching arrangements in order to better characterize the conditions that encourage maximal professional development for the instructors involved.

### Acknowledgements
We thank Warren Code, Jessica Dawson, Joss Ives, Ido Roll, Sarah Bean Sherman, the UBC SES community, the PHASER group, and the Instructors A, B, C, D, Y, and Z. This extension of CWSEI work is funded by John and Deb Harris, the UBC Faculty of Science, and the UBC Department of Physics and Astronomy. The study was approved by the UBC Behavioural Research Ethics Board (H14-01879).

### References
[1] Freeman, S., Eddy, S. L., McDonough, M., Smith, M. K., Okoroafor, N., Jordt, H., & Wenderoth, M. P. (2014). Active learning increases student performance in science, engineering, and mathematics. *Proceedings of the National Academy of Sciences*, 111(23), 8410-8415.

[2] Henderson, C., & Dancy, M. H. (2009). Impact of physics education research on the teaching of introductory quantitative physics in the United States. *Physical Review Special Topics Physics Education Research*, 5(2), 020107.

[3] Henderson, C., Dancy, M., & Niewiadomska-Bugaj, M. (2012). Use of research-based instructional strategies in introductory physics: Where do faculty leave the innovation-decision process? *Physical Review Special Topics Physics Education Research*, 8(2), 020104.

[4] Henderson, C., Beach, A., & Famiano, M.. (2009). Promoting instructional change via co-teaching. *American Journal of Physics (Physics Education Research Section)*, 77(3), 274-283.

[5] Henderson, C., Beach, A., & Finkelstein, N.. (2011). Facilitating change in undergraduate STEM instructional practices: An analytic review of the literature. *Journal of Research in Science Teaching*, 48(8), 952-984.

[6] Carl Wieman Science Education Initiative at the University of British Columbia. (2015, June). Retrieved from http://www.cwsei.ubc.ca

[7] Wieman, C., Perkins, K., & Gilbert, S.. (2010). Transforming science education at large research universities: A case study in progress. *Change: The Magazine of Higher Learning*, 42, 6-14.

[8] Wieman, C., Deslauriers, L., & Gilley, B.. (2013). Use of research-based instructional strategies: How to avoid faculty quitting. *Physical Review Special Topics Physics Education Research*, 9(2), 023102.

[9] Gibbs, G.R.. (2008). *Analysing qualitative data*. Thousand Oaks, CA, USA: SAGE Publications USA.

[10] Wieman, C. & Gilbert, S.. (2014). The teaching practices inventory: A new tool for characterizing college and university teaching in mathematics and science. *CBE-Life Sciences Education*, 13(3), 552-569.

[11] Smith, M.K., Jones, F.H.M., Gilbert, S.L., & Wieman, C.E.. (2013). The Classroom Observation Protocol for Undergraduate STEM (COPUS): A new instrument to characterize university STEM classroom practices. *CBE-Life Sciences Education*, 12(4), 618-627.

[12] Yin, R. K. (2013). *Case study research: Design and methods*. Sage publications USA.

[13] Lane, E.S. & Harris, S.E.. (2015). A new tool for measuring student behavioral engagement in large university classes. *Journal of College Science Teaching*, 44(6), 83-91.

[14] Collins, A., Brown, J.S., & Holum, A.. (1991). Cognitive apprenticeship: Making thinking visible. *American educator*, 15(3), 6-11.

[15] Heiner, C.E., Banet, A.I., & Wieman, C.. (2014). Preparing students for class: How to get 80% of students reading the textbook before class. *American Journal of Physics*, 82(10), 989-996.

[16] Jones, D.J., Madison, K.W., & Wieman, C.E.. (2015). Transforming a 4th year modern optics course using a deliberate practice framework. *Physical Review Special Topics Physics Education Research*, 11(2), 020108.

| Novice instructor | A | B | C | D |
|---|---|---|---|---|
| Course (year) | 1 (2013) | 1 (2014) | 1 (2015) | 2 (2015) |
| Course context* | First-year large-scale calculus based course using active learning techniques. Multiple sections and instructors. Structure and materials established. | | | |
| Prior teaching experience of novice* | <1 year teaching. No experience with RBIS. | <5 years teaching. Some previous exposure to RBIS through the CWSEI. | 10 years teaching at all levels. Some previous exposure to RBIS through the CWSEI. | 10 years teaching at all levels. Some previous exposure to RBIS through the CWSEI. |
| Novice's position | Research stream tenure-track. | Teaching stream contract. | Research stream tenured. | Research stream tenured. |
| Approach of novice* | Intention to learn "tried and tested" methods. | Saw paired teaching as an "apprenticeship." | Sought feedback from expert, but "most of the things weren't new." | Focused on in-class product and not professional development. |
| Expert instructor | Instructor Y, teaching stream tenured, 20 years teaching experience, 10 years PER experience. | | | Instructor Z, teaching stream tenure-track, 20 years teaching experience, 10 years PER experience. |
| Relationship with expert* | "Incredibly friendly." | "… I do like them as [a person]." | "It was very collegial." | "… we all got along." |
| Instructor meetings | No dedicated meetings between paired instructors. Informal meetings before and after class and email communication. Weekly whole-team instructor meetings. | | | |
| Teaching assignment sequence* | Taught course 1 individually in next two years. | Taught both course 1 and other similar courses in subsequent year. | Taught junior level course 3 at the same time as pair-teaching. Next year will teach courses 1 and 3 individually. | Will teach course 2 individually next year. |

| **SES support*** | No SES support. | SES provided feedback based on classroom observations and informal student interviews. |

Table I: Factors characterizing the four paired teaching arrangements. * = Suggested connections between these factors and learning about teaching are discussed in section III C. SES = science education specialist.

## Appendix A. Interview Protocol for Paired Instructors

Here we list the interview questions that were asked to instructors shortly following their paired teaching experience.

1) What concerns arose during the term and how did you and the team work through those?

2) What turned out to be the biggest challenge for you?

3) What turned out to be the biggest benefit to you?

4) What was the biggest surprise for you?

5) How did the time commitment compare to teaching it by yourself?

6) What kind of interactions did you have with your co-instructor? Which were most useful? Least useful?

7) What kind of interactions with the science education specialist were most useful? Least useful?

8) How do you think the co-teaching program could be improved?

9) What worked? What didn't work for you?

10) What did you learn about teaching?

11) As a result of this experience, did your teaching philosophy change? If so how and why?

12) What teaching methods used in this experience do you anticipate using again in the future?